\documentclass[12pt]{article}
\usepackage{graphicx}
\usepackage[utf8]{inputenc}
\usepackage[english
]{babel}
\def\baselinestretch{1.5}
\begin{document}
\begin{center}
\bf{God plays coins or superposition principle for classical probabilities in quantum suprematism representation of qubit states.}\\
\end{center}
\bigskip

\begin{center} {\bf V. N. Chernega$^1$, O. V. Man'ko$^{1,2}$, V. I. Man'ko$^{1,3}$}
\end{center}

\medskip

\begin{center}
$^1$ - {\it Lebedev Physical Institute, Russian Academy of Sciences\\
Leninskii Prospect 53, Moscow 119991, Russia}\\
$^2$ - {\it Bauman Moscow State Technical University\\
The 2nd Baumanskaya Str. 5, Moscow 105005, Russia}\\
$^3$ - {\it Moscow Institute of Physics and Technology (State University)\\
Institutskii per. 9, Dolgoprudnyi, Moscow Region 141700, Russia}\\
Corresponding author e-mail: manko@sci.lebedev.ru
\end{center}

\section*{Abstract}
For three given probability distributions describing positions of three classical coins the quantum density matrix of spin-$1/2$ state is constructed and its matrix elements are associated with triada of Malevich's squares. The superposition principle of spin-$1/2$ states is presented in the form of nonlinear addition rule for these classical coin probabilities. We illustrate the formulas by the statement"God does not play dice - God plays coins".

\section{Introduction}
The states of a quantum system determined by the vectors $|\psi\rangle$ in a Hilbert space \cite{Dirac} have the property that the superpositions of the vectors $c_1|\psi_1\rangle+c_2|\psi_2\rangle$ with complex coefficients $c_1$ and $c_2$ always describe the other states of the system. The superposition of the vectors corresponds to superposition of wave functions $c_1\psi_1(q)+c_2\psi_2(q)$ describing the states of the systems \cite{Sch26}. For density operators $\hat\rho_\psi=|\psi\rangle\langle\psi|$  \cite{Landau27,vonNeumann27} the explicit formula for the superposition of the state $\hat\rho_1=|\psi_1\rangle\langle\psi_1|$ and $\hat\rho_2=|\psi_2\rangle\langle\psi_2|$ was obtained in \cite{Sudarshan1,Sudarshan2}. For normalized pure states the density operators $\hat\rho_\psi$, $\hat\rho_1$, $\hat\rho_2$ have the geometrical meaning of the projectors. The superposition principle and its representation in the form of the addition rule for two projectors give the answer to the following question: How to add two projectors in order to obtain another projector? Recently the tomographic probability representation of quantum states was introduced both for continious variables \cite{TombesiPLA} and discrete spin variables \cite{DodPLA,OlgaJETP,Bregence,Weigert1,Weigert2,Paini} (see also review \cite{IbortPhysScr150}). In this representation the quantum system states are described by fair probability distributions. For example, the states of spin-$1/2$ particle associated with state density $2\times2$-matrix can be determined by three probability distributions (of dichotomic variables) \cite{Marmo} which are illustrated in quantum suprematism representation by triada of Malevich's squares \cite{Chernega1,Chernega2,Chernega3}.

In this connection we consider the following problem. Given three probability distributions of three classical dichotomic random variables. We can interpret these probability distributions as probability distributions describing position "up" or "down" of three classical coins. Using the interpretation of the quantum states of the spin-$1/2$ particle in terms of the three probability distributions \cite{Marmo,Chernega1,Chernega2,Chernega3} we have possibility to use the superposition principle known for density operators of the states \cite{Sudarshan1,Sudarshan2} to obtain the rule of addition of the classical coin probabilities corresponding to the quantum state superposition.

The aim of our work is to study the bijective map of the three classical coin positions onto quantum spin-$1/2$ state density matrix and to give explicit formula for nonlinear addition rule of the probabilities determining positions of the classical coins corresponding to linear superposition of the state vectors in the Hilbert space.

The paper is organized as follows. In Sec.2 the review of superposition principle for density operators of pure quantum states \cite{Sudarshan1,Sudarshan2} is given. In Sec.3 the suprematism representation of qubit states \cite{Chernega1,Chernega2,Chernega3} is discussed. In Sec.4 the addition rule for classical probabilities describing three coin positions is derived using the superposition principle for corresponding orthogonal quantum states. In Sec. 5 the case of nonorthogonal states is studied. The conclusions and prospectives are given in Sec.6.

\section{Superposition of quantum states}
Let us take density operators $\hat\rho_1$ and $\hat\rho_2$ of the pure states determined by the vectors $|\psi_1\rangle$ and $|\psi_2\rangle$, respectively, satisfying the normalisation and orthogonality conditions
$\langle\psi_1|\psi_1\rangle=\langle\psi_2|\psi_2\rangle=1, \quad \langle\psi_1|\psi_2\rangle=0.$
Thus we have $
\hat\rho_1^2=\hat\rho_1, \quad \hat\rho_2^2=\hat\rho_2, \quad \hat\rho_1\hat\rho_2=0.$
The pure state
\begin{equation}\label{eq.3}
|\psi\rangle=c_1|\psi_1\rangle+c_2|\psi_2\rangle
\end{equation}
is normalized pure state if
\begin{equation}\label{eq.4}
|c_1|^2+|c_2|^2=1
\end{equation}
One can take the coefficients $c_1$ and $c_2$ satisfying the relations
\[c_1^\ast=c_1,\quad c_2=|c_2|\exp(i\phi).\]
The density operator of the pure state (\ref{eq.3}) has the form
\begin{equation}\label{eq.5}
\hat\rho_\psi=|\psi\rangle\langle\psi|=c_1^2|\psi_1\rangle\langle\psi_1|+|c_2|^2|\psi_2\rangle\langle\psi_2|+
c_1|c_2|e^{-i\phi}|\psi_1\rangle\langle\psi_2|+c_1|c_2|e^{i\phi}|\psi_2\rangle\langle\psi_1|.
\end{equation}
The operator $\hat\rho_\psi$ determines the pure state and satisfies the condition $\hat\rho_\psi^2=\hat\rho_\psi,\,\mbox{Tr}\hat\rho^2_\psi=1$. The parameters $c_1$ and $|c_2|$ determine two probabilities $\lambda_1=c_1^2,$ $\lambda_2=|c_2|^2,$ $\lambda_1+\lambda_2=1$ and the phase parameter $\phi$ corresponds to the relative phase for addition of two wave functions responsible for quantum interference phenomenon. The formula (\ref{eq.5}) was presented in the form of adding two orthogonal projectors \cite{Sudarshan1,Sudarshan2}, i.e.
\begin{equation}\label{eq.6}
\hat\rho_\psi=\lambda_1\hat\rho_1+\lambda_2\hat\rho_2+
\sqrt{\lambda_1\lambda_2}\,\,\frac{\hat\rho_1\hat\rho_0\hat\rho_2+\hat\rho_2\hat\rho_0\hat\rho_1}{\sqrt{\mbox{Tr}(\hat\rho_1\hat\rho_0\hat\rho_2\hat\rho_0)}}.
\end{equation}
Here $\hat\rho_1=|\psi_1\rangle\langle\psi_1|,\,\hat\rho_2=|\psi_2\rangle\langle\psi_2|$ and $\hat\rho_0=|\psi_0\rangle\langle\psi_0|,$ where
\begin{equation}\label{eq.7}
\langle\psi_0|\psi_0\rangle=1, \quad \langle\psi_1|\psi_0\rangle=e^{i\phi_1}, \quad \langle\psi_2|\psi_0\rangle=e^{i\phi_2}
\end{equation}
and the relative phase $\phi=\phi_2-\phi_1$. We can take $\phi_1=0$. Thus the arbitrary vector $|\psi_0\rangle$ is used to determine the arbitrary relative phase $\phi$ corresponding to superposition of vectors $|\psi_1\rangle$ and $|\psi_2\rangle$ in formulae (\ref{eq.3}). One can check that the Hermitian operator $\hat\rho_\psi$ given by (\ref{eq.6}) satisfies the conditions $\hat\rho_\psi^\dagger=\hat\rho_\psi,\, \mbox{Tr}\hat\rho_\psi=1,\,\hat\rho^2_\psi=\hat\rho_\psi$. Due to the matrix relation (\ref{eq.6}) all the matrix elements of the matrix $\hat\rho_\psi$ are equal to matrix elements of the matrix in right-hand side of the equality.

\section{Spin-$1/2$ states in terms of probabilities of three coin positions and quantum suprematism representation.}
In \cite{Chernega1,Chernega2,Chernega3} it was demonstrated that a density matrix of spin-$1/2$ system can be expressed in terms of the probabilities $0\leq p_1,p_2,p_3\leq 1$ of three classical coin positions. The construction of the density matrix is presented as follows. Given three probability distributions defined by three probability vectors
\begin{equation}\label{eq.8}
\vec{\cal P}_1=\left(\begin{array}{c}
p_{1}\\
1-p_{1}\end{array}\right),\quad
\vec{\cal P}_2=\left(\begin{array}{c}
p_{2}\\
1-p_{2}\end{array}\right),\quad
\vec{\cal P}_3=\left(\begin{array}{c}
p_{3}\\
1-p_{3}\end{array}\right).
\end{equation}
The numbers $p_1,\,p_2,\,p_3$ can be interpreted as the probabilities to obtain for three classical coins the positions "up". The numbers $1-p_1,\,1-p_2,\,1-p_3$ are interpreted as the probabilities to obtain for these coins the positions "down". Thus, the random positions of three coins "up" or "down" are associated with the probability vectors (\ref{eq.8}). One can consider \cite{Chernega3} the classical random variables determined by the vectors
\[\vec X=\left(\begin{array}{c}
x_{1}\\
x_{2}\end{array}\right),\quad \vec Y=\left(\begin{array}{c}
y_{1}\\
y_{2}\end{array}\right),\quad \vec Z=\left(\begin{array}{c}
z_{1}\\
z_{2}\end{array}\right)\]
with real components given in the form of 
vectors (\ref{eq.8}). One can choose \cite{Chernega2,Chernega3} the classical random variables determined by the vectors with components
\begin{equation}\label{eq.9}
\vec X=\left(\begin{array}{c}
x\\
-x\end{array}\right),\quad \vec Y=\left(\begin{array}{c}
y\\
-y\end{array}\right),\quad \vec Z=\left(\begin{array}{c}
z_{1}\\
z_{2}\end{array}\right)
\end{equation}
and associate with first coin positions vector $\vec X$, second coin positions vector $\vec Y$ and third coin positions vector $\vec Z$. The statistics of the dichotomic random variables is given by the moments determined by the probability vectors. For example, mean values of the random variables read
\begin{eqnarray}
&&\langle\vec X\rangle=p_1 x -(1-p_1)x=\vec X\vec{\cal P}_1,\nonumber\\
&&\langle\vec Y\rangle=p_2 y -(1-p_2)y=\vec Y\vec{\cal P}_2,\label{eq.10}\\
&&\langle\vec Z\rangle=p_3 z_1+(1-p_3)z_2=\vec Z\vec{\cal P}_3.\nonumber
\end{eqnarray}
The second moments are given by the formulas
\begin{eqnarray}
&&\langle\vec X^2\rangle=x^2[p_1+(1-p_1)]=x^2,\nonumber\\
&&\langle\vec Y^2\rangle=y^2[p_2 +(1-p_2)]=y^2,\label{eq.11}\\
&&\langle\vec Z^2\rangle=z_1^2p_3+z_2^2(1-p_3)=(z_1^2-z_2^2)p_3+z_2^2.\nonumber
\end{eqnarray}
In \cite{Chernega2,Chernega3} it was suggested to construct the quantum mechanics of spin-$1/2$ system using the model of three classical probabilities and three classical random variables given by the vectors $\vec X,\,\vec Y,\,\vec Z$. To present this model we construct two matrices: first one with matrix elements expressed in terms of probabilities $p_1,p_2,p_3$, i.e.
\begin{equation}\label{eq.12}
\rho=\left(\begin{array}{cc}
p_3&(p_1-1/2)-i(p_2-1/2)\\
p_1-1/2+i(p_2-1/2)&1-p_3
\end{array}\right)
\end{equation}
and second one with matrix elements expressed in terms of $x,\,y,\,z_1,\,z_2$, i.e.
\begin{equation}\label{eq.13}
H=\left(\begin{array}{cc}
z_1&x-i y\\
x+i y&z_2
\end{array}\right).
\end{equation}
One can see that by construction the matrices $\rho$ and $H$ are Hermitian matrices, i.e. $\rho^\dag=\rho,\,H^\dag=H$. Also the matrix $\rho$ has the unit trace, i.e. $\mbox{Tr}\rho=1$. In our model of three classical coins there is no correlations among the coins.
It means that all the probabilities of coin positions belong to domains $0\leq p_1\leq 1,\,0\leq p_2\leq1,\,0\leq p_3\leq 1$. Let us impose the condition corresponding to a specific correlations in the coin behaviour, namely that the numbers $p_1,\,p_2,\,p_3$ satisfy the inequality
\begin{equation}\label{eq.14}
(p_1-1/2)^2+(p_2-1/2)^2+(p_3-1/2)^2\leq 1/4.
\end{equation}
This inequality means that the classical coin positions are correlated. Also the inequality (\ref{eq.14}) provides the condition for Hermitian matrix (\ref{eq.12}) to have only nonnegative eigenvalues, i.e. $\rho\geq0$. Such Hermitian matrix $\rho$ with unit trace and nonnegative eigenvalues in quantum mechanics is the density matrix of spin-$1/2$ system. On the other hand in quantum mechanics the Hermitian matrix (\ref{eq.13}) corresponds to an observable. The statistical properties of this observable can be expressed in terms of properties of classical random variables $\vec X,\,\vec Y,\,\vec Z$, e.g.the quantum mean value of the observable $H$ is
\begin{equation}\label{eq.15}
\langle H\rangle=\mbox{Tr}(\rho H)=\langle\vec X\rangle+\langle \vec Y\rangle+\langle\vec Z\rangle,
\end{equation}
where we use the classical random variable means (\ref{eq.10}).

The probabilities $p_1,\,p_2,\,p_3$ can be used to construct three Malevich's squares\\
(Figure.1) \cite{Chernega1,Chernega2,Chernega3} with the sides
\begin{eqnarray}
&&L_1=\left(2+2p_1^2-4p_1-2p_{2}+2p_{2}^2+2p_1 p_{2}\right)^{1/2},\nonumber\\
&&L_2=\left(2+2p_2^2-4p_2-2p_{3}+2p_{3}^2+2p_2 p_{3}\right)^{1/2},\nonumber\\
&&L_3=\left(2+2p_3^2-4p_3-2p_{1}+2p_{1}^2+2p_3 p_{1}\right)^{1/2}.\label{eq.M2}
\end{eqnarray}

Thus, the classical probability distributions are associated with triada of Malevich's black, red and white squares \cite{Chernega1} where the numbers $p_1,\,p_2,\,p_3$  must satisfy the inequality (\ref{eq.14}). The approach called quantum suprematism representation of spin-$1/2$ states provides the possibility to map the density matrix of qubit states onto triada of Malevich's squares on the plane. For classical coins the sides of Malevich squares are functions of arbitrary probabilities $0\leq p_1,p_2,p_3\leq1$. For spin-$1/2$ states the probabilities satisfy inequality (\ref{eq.14}).
\begin{figure}
  \centering
  \includegraphics[width=70mm]{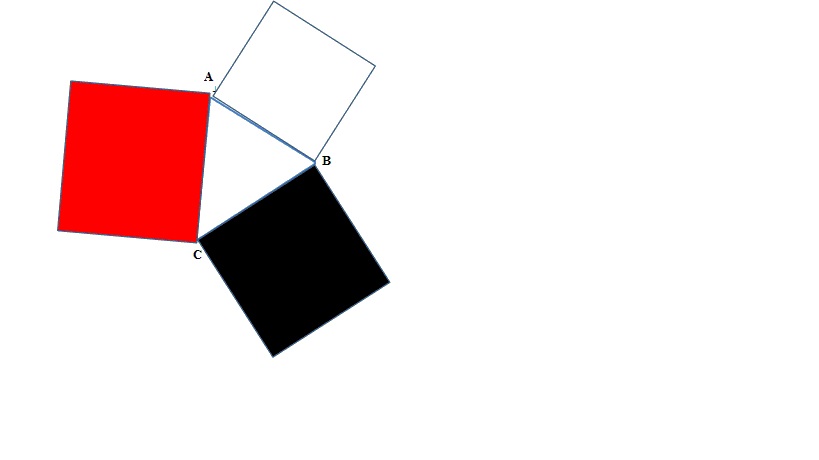}\\
  \caption{Triada of Malevich's squares}\label{1}
\end{figure}

\section{Superposition principle}
In this section we present the superposition principle of quantum states of spin-$1/2$ system in quantum suprematism representation. This means that we give the addition rule for pairs of triadas of Malevich's squares. As the result of this addition we obtain another triada of Malevich's squares. All triadas correspond to pure quantum states $\hat\rho_\psi,\, \hat\rho_1,\,\hat\rho_2$. In fact, we have to calculate the probabilities $p_1^{(\psi)},\,p_2^{(\psi)},\, p_3^{(\psi)}$ providing the density matrix corresponding to operator of pure state $\hat\rho_\psi$ if the probabilities $p_1,\,p_2,\,p_3$ determining the pure state $\hat\rho_1=|\psi_1\rangle\langle\psi_1|$ and probabilities ${\cal P}_1,\,{\cal P}_2,\,{\cal P}_3$ determining the pure state $\hat\rho_2=|\psi_2\rangle\langle \psi_2|$ are given.

Thus we have the four density matrices of the form (\ref{eq.12}). These matrices correspond to operators
$\hat\rho_1=|\psi_1\rangle\langle\psi_1|,$ $\hat\rho_2=|\psi_2\rangle\langle\psi_2|,$  $\hat\rho_0=|\psi_0\rangle\langle\psi_0|,$
$\hat\rho_\psi=|\psi\rangle\langle\psi|.$
We introduce the notations
$p=p_1-1/2-i(p_2-1/2,$ ${\cal P}={\cal P}_1-1/2-i({\cal P}_2-1/2),$ $\Pi=\Pi_1-1/2-i(\Pi_2-1/2),$  $p^{(\psi)}=p^{(\psi)}_1-1/2-i(p^{(\psi)}_2-1/2)$ and the four matrices corresponding to four operators 
can be expressed in terms of the probabilities. For example
\begin{eqnarray}\label{eq.18}
&&\rho_1=\left(\begin{array}{cc}
p_3&p\\
p^\ast&1-p_3
\end{array}\right).
\end{eqnarray}
Since the state $\hat\rho_1=|\psi_1\rangle\langle\psi_1|$ is pure state the probabilities $(p_1,\,p_2,\,p_3)=\vec p$ which are components of the vector $\vec p$ satisfy the equality
\begin{equation}\label{eq.18a}
(p_1-1/2)^2+(p_2-1/2)^2+(p_3-1/2)^2= 1/4.
\end{equation}
Analogously the density matrix of the state $\hat\rho_2=|\psi_2\rangle\langle\psi_2|$ is expressed in terms of probabilities ${\cal P}_1,\,{\cal P}_2,\,{\cal P}_3$ as
\begin{equation}\label{eq.19}
\rho_2=\left(\begin{array}{cc}
{\cal P}_3&{\cal P} \\
{\cal P}^\ast&1-{\cal P}_3
\end{array}\right),\quad
{\cal P}={\cal P}_1-1/2-i({\cal P}_2-1/2).
\end{equation}
For pure states $\rho_2$ the components ${\cal P}_1,\,{\cal P}_2,\,{\cal P}_3$ of the vector $\vec{\cal P}$ satisfy the condition (\ref{eq.18a}).\\
Fidelity for two states $\rho_1,$ $\rho_2$ reads
\begin{equation}\label{eq.18b}
{\cal F}=\mbox{Tr}\,(\rho_1\rho_2)=2+2(\vec p\cdot\vec{\cal P})-p_1-p_2-p_3-{\cal P}_1-{\cal P}_2-{\cal P}_3.
\end{equation}
Purity of the state $\rho$ in terms of probabilities reads
\begin{equation}\label{eq.18c}
\mu=\mbox{Tr}\,\rho^2=2\left(1+\vec p^2-p_1-p_2-p_3\right).
\end{equation}
The density operator of the pure state $\hat\rho_0=|\psi_0\rangle\langle\psi_0|$ which is the projector operator providing the relative phase parameter depends on three probabilities
\begin{equation}\label{eq.20}
\rho_0=\left(\begin{array}{cc}
\Pi_3&\Pi\\
\Pi^\ast &1-\Pi_3
\end{array}\right), \quad\Pi=\Pi_1-1/2-i(\Pi_2-1/2).
\end{equation}
The probabilities $\Pi_1,\,\Pi_2,\,\Pi_3$ satisfy the equality (\ref{eq.18a}). The density operator of the superposition state $\hat\rho_\psi=|\psi\rangle\langle\psi|$ has the density matrix
\begin{equation}\label{eq.21}
\rho_\psi=\left(\begin{array}{cc}
p^{(\psi)}_3&p^{(\psi)}\\
(p^{(\psi)})^\ast&1-p^{(\psi)}_3
\end{array}\right),\quad p^{(\psi)}=p^{(\psi)}_1-1/2-i(p^{(\psi)}_2-1/2)
\end{equation}
and the probabilities $p^{(\psi)}_1,\,p^{(\psi)}_2,\,p^{(\psi)}_3$ must satisfy the equality (\ref{eq.18a}).

The superposition state formula (\ref{eq.6}) can be presented as result of addition of the probabilities given in the form of $3$-vectors
\begin{equation}\label{eq.22}
\vec P=\left(\begin{array}{c}
p_1\\p_2\\p_3\end{array}\right),\quad
\vec{\cal P}=\left(\begin{array}{c}
{\cal P}_1\\{\cal P}_2\\{\cal P}_3\end{array}\right),\quad
\vec \Pi=\left(\begin{array}{c}
\Pi_1\\\Pi_2\\\Pi_3\end{array}\right),\quad
\vec P_\psi=\left(\begin{array}{c}
p^{(\psi)}_1\\p^{(\psi)}_2\\p^{(\psi)}_3\end{array}\right).
\end{equation}
The result of addition given as vector $\vec P_\psi$ has two contributions
\begin{equation}\label{eq.23}
\vec P_\psi=(\lambda_1\vec P+\lambda_2\vec{\cal P})+\sqrt{\lambda_1\lambda_2}\vec\Delta.
\end{equation}
The vector $\vec\Delta$ nonlinearly depends on components of vectors $\vec P,\,\vec{\cal P},\,\vec\Pi$ and the nonlinear contribution is responsible for the equality
$\rho^2_\psi=\rho_\psi.$
To give the explicit form of the components of the vector $\vec \Delta$ we introduce the notations for $\mbox{Tr}(\rho_1\rho_0\rho_2\rho_0)=T^2$ and get the expression of this factor $T$ in terms of the probabilities. It reads
\begin{eqnarray}
T=&&\left\{\left[\left(p_3\pi_3+p\Pi^\ast\right){\cal P}_3+\left(p_3\pi+p(1-\Pi_3)\right){\cal P}^\ast\right]\Pi_3+\left[\left( p^\ast\Pi_3+(1-p_3)\Pi^\ast\right){\cal P}+\right.\right.\nonumber\\
&&\left.\left(p^\ast\Pi+(1-p_3)(1-\Pi_3)\right)\right](1-\Pi_3)+
\left[\left(p^\ast\Pi_3+(1-p_3)\Pi^\ast\right){\cal P}_3+\right.\nonumber\\
&&\left.\left( p^\ast\Pi+(1-p_3)(1-\Pi_3)\right){\cal P}^\ast\right]\Pi^\ast+\left[\left(p_3\Pi_3+p\Pi^\ast\right){\cal P}\right.+\nonumber\\
&&\left.\left.\left(p_3\Pi+p(1-\Pi_3)\right)(1-{\cal P}_3)\right]\Pi\right\}^{-1/2}.\label{eq.25}
\end{eqnarray}
The components of vector $\vec\Delta$ read
\begin{eqnarray}
\Delta_3=&&\left\{\left(p_3\Pi_3+p\Pi^\ast\right){\cal P}_3+\left[p_3\Pi+p(1-\Pi_3)\right]{\cal P}^\ast
+({\cal P}_3\Pi_3+{\cal P}\Pi^\ast)p_3+\right.\nonumber\\
&&\left.\left[{\cal P}_3\Pi+{\cal P}(1-\Pi_3)\right]p^\ast\right\}T,\label{eq.26}
\end{eqnarray}
\begin{eqnarray}
\Delta_1=&&\left\{\frac{1}{2}+\mbox{Re}\left\{\left[p^\ast\Pi_3+(1-p_3)\Pi^\ast\right]{\cal P}_3+\left[p^\ast\Pi+(1-p_3)(1-\Pi_3)\right]{\cal P}^\ast+\right.\right.\nonumber\\
&&\left.\left.\left[{\cal P}^\ast\Pi_3+(1-{\cal P}_3)\Pi^\ast\right]p_3+\left[{\cal P}^\ast\Pi+(1-{\cal P}_3)(1-\Pi_3)\right]p^\ast\right\}\right\}T,\label{eq.27}
\end{eqnarray}
\begin{eqnarray}
\Delta_2=&&\left\{\frac{1}{2}+\mbox{Im}\left\{\left[p^\ast\Pi_3+(1-p_3)\Pi^\ast\right]{\cal P}_3+\left[p^\ast\Pi+(1-p_3)(1-\Pi_3)\right]{\cal P}^\ast+\right.\right.\nonumber\\
&&\left.\left.\left[{\cal P}^\ast\Pi_3+(1-{\cal P}_3)\Pi^\ast\right]p_3+\left[{\cal P}^\ast\Pi+(1-{\cal P}_3)(1-\Pi_3)\right]p^\ast\right\}\right\}T,\label{eq.28}
\end{eqnarray}
The formula (\ref{eq.23}) with factor $T$ (\ref{eq.25}) and components of vector $\vec \Delta$ (\ref{eq.26})-(\ref{eq.28}) gives the nonlinear addition rule for probabilities $p_1,\,p_2,\,p_3$ and ${\cal P}_1,\,{\cal P}_2,\,{\cal P}_3$. This formula corresponds to superposition of two orthogonal states $|\psi_1\rangle$ and $|\psi_2\rangle$.

We give another derivation of the addition rule for the probabilities $p_j,\,{\cal P}_j$ which follows from the superposition formula $|\psi\rangle=c_1|\psi_1\rangle+c_2|\psi_2\rangle$ where $\langle\psi_1|\psi_2\rangle=0$. To do this we observe that arbitrary Pauli spinor $|\chi\rangle=\left(\begin{array}{c}
\chi_1\\\chi_2\end{array}\right)$ can be expressed in terms of probabilities. Namely let $\chi_1=e^{i\alpha}a,\,\chi_2=e^{i\beta}b$, and here $a>0,\,b>0$. Thus
\begin{equation}\label{eq.29}
|\chi\rangle=e^{i\alpha}\left(\begin{array}{c}
a\\b e^{i\gamma}\end{array}\right),\quad \gamma=(\beta-\alpha).
\end{equation}
Using probabilities $p_1,\,p_2,\,p_3$ the Pauli spinor can be given in the form
\begin{equation}\label{eq.30}
|\chi\rangle=e^{i\alpha}\left(\begin{array}{c}
\sqrt{p_3}\\\sqrt{1-p_3}e^{i\gamma}\end{array}\right),
\end{equation}
where we introduce the probabilities in view of relations
\begin{equation}\label{eq.31}
a^2=p_3,\quad b^2=1-p_3,\quad \cos\gamma=\frac{p_1-1/2}{\sqrt{p_3(1-p_3)}},\quad\sin\gamma=\frac{p_2-1/2}{\sqrt{p_3(1-p_3)}}.
\end{equation}
The phase $\alpha$ is an arbitrary angle parameter. The value of this parameter does not change density operator of pure state $|\psi\rangle\langle \psi|=\hat \rho_\psi$ which is gauge invariant projector operator.

Let us consider four vectors $|\psi_1\rangle,\,|\psi_2\rangle,\,|\psi_0\rangle,\,|\psi\rangle,$ and introduce the notations
\begin{equation}\label{eq.32}
|\psi_1\rangle=e^{-i\alpha}|\chi\rangle=\left(\begin{array}{c}
a\\b e^{i\beta}\end{array}\right)=\left(\begin{array}{c}
\sqrt{p_3}\\\sqrt{1-p_3}e^{i\beta}\end{array}\right).
\end{equation}
In (\ref{eq.32})
\begin{equation}\label{eq.33}
\cos\beta=\frac{p_1-1/2}{\sqrt{p_3(1-p_3)}},\quad \sin\beta=\frac{p_2-1/2}{\sqrt{p_3(1-p_3)}}.
\end{equation}
The unit vector $|\psi_2\rangle$ which we take to be orthogonal to vector $|\psi_1\rangle$ has the form
\begin{equation}\label{eq.34}
|\psi_2\rangle=\left(\begin{array}{c}
\sqrt{{\cal P}_3}\\\sqrt{1-{\cal P}_3}e^{i\mu}\end{array}\right),\quad 1>{\cal P}_3>0,\quad\mu=\beta\pm\pi/2,
\end{equation}
where
\begin{equation}\label{eq.35}
\cos\mu=\frac{{\cal P}_1-1/2}{\sqrt{{\cal P}_3(1-{\cal P}_3)}},\quad \sin\mu=\frac{{\cal P}_2-1/2}{\sqrt{{\cal P}_3(1-{\cal P}_3)}}.
\end{equation}
The unit vector $|\psi_0\rangle$ is of the form
\begin{equation}\label{eq.36}
|\psi_0\rangle=\left(\begin{array}{c}
\sqrt{\Pi_3}\\\sqrt{1-\Pi_3}e^{i\delta}\end{array}\right),
\end{equation}
where
\begin{equation}\label{eq.37}
\cos\delta=\frac{\Pi_1-1/2}{\sqrt{\Pi_3(1-\Pi_3)}},\quad \sin\delta=\frac{\Pi_2-1/2}{\sqrt{\Pi_3(1-\Pi_3)}}.
\end{equation}
Superposition vector $|\psi\rangle$ is defined as
\begin{equation}\label{eq.38}
|\psi\rangle=\sqrt{\Pi_3}|\psi_1\rangle+\sqrt{1-\Pi_3}e^{i\delta}|\psi_2\rangle.
\end{equation}
It is given by the column vector
\begin{equation}\label{eq.39}
|\psi\rangle=\left(\begin{array}{c}
\sqrt{\Pi_3p_3}+e^{i\delta}\sqrt{{\cal P}_3(1-\Pi_3)}\\
e^{i\beta}\sqrt{\Pi_3(1-p_3)}+e^{i(\delta+\mu)}\sqrt{(1-\Pi_3)(1-{\cal P}_3)}
\end{array}\right).
\end{equation}
Calculating the matrix $|\psi\rangle\langle\psi|$ we reproduce result corresponding to the matrix form of superposition principle given by (\ref{eq.6}).

\section{Derivation of probability representation for\\ superposition-state expression in case \\ of arbitrary pure states}
In this Section for one qubit we provide the formula for density matrix of superposition of two arbitrary pure states. The first state is given by the vector $|\psi_1\rangle$ of the form (\ref{eq.32}) where $\beta=\phi_1$, i.e.
\begin{equation}\label{S1}
|\psi_1\rangle=\left(\begin{array}{c}
\sqrt{p_3}\\
\sqrt{1-p_3}e^{i\phi_1}
\end{array}\right).
\end{equation}
Here $p_3$ is probability of spin-projection $m=+1/2$ on the $z$-direction. Since
\begin{equation}\label{S2}
\langle \psi_1|=\left(\sqrt{p_3},\,\sqrt{1-p_3}e^{-i\phi_1}\right)
\end{equation}
the explicit density matrix $\rho=|\psi_1\rangle\langle \psi_1|$ of this pure state expressed in terms of probabilities $p_1,\,p_2,\,p_3$ reads
\begin{equation}\label{S3}
|\psi_1\rangle\langle \psi_1|=\left(\begin{array}{cc}
\sqrt{p_3}&
\sqrt{p_3(1-p_3)}e^{-i\phi_1}\\
\sqrt{p_3(1-p_3)}e^{i\phi_1}&1-p_3
\end{array}\right).
\end{equation}
It means that the phase $\phi_1$ in (\ref{S1}) is determined by the probabilities $p_1$ and $p_2$ of spin-$1/2$ projections $m=+1/2$ on the axes $x$ and $y$, respectively, i.e.
\begin{equation}\label{S4}
\sqrt{p_3(1-p_3)}e^{-i\phi_1}=p_1-1/2-i(p_2-1/2).
\end{equation}
This equality gives the relations for the phase $\phi_1$ of the form
\begin{equation}\label{S5}
\cos\phi_1=\frac{p_1-1/2}{\sqrt{p_3(1-p_3)}},\quad \sin\phi_1=\frac{p_2-1/2}{\sqrt{p_3(1-p_3)}}.
\end{equation}
Let us introduce the second state-vector $|\psi_2\rangle$ of the qubit expressed in terms of three probabilities ${\cal P}_1,\,{\cal P}_2,\,{\cal P}_3$ of the spin-$1/2$ projections $m=+1/2$ on the axes $x,\,y,\,z$, respectively, as
\begin{equation}\label{S6}
|\psi_2\rangle=\left(\begin{array}{c}
\sqrt{{\cal P}_3}\\
\sqrt{1-{\cal P}_3}e^{i\phi_2}\end{array}\right).
\end{equation}
The phase $\phi_2$ is not related with phase $\phi_1$. Analogously to (\ref{S5}) one has the relation of the phase $\phi_2$ with probabilities ${\cal P}_1$ and ${\cal P}_2$ of the form
\begin{equation}\label{S7}
\cos\phi_2=\frac{{\cal P}_1-1/2}{\sqrt{{\cal P}_3(1-{\cal P}_3)}},\quad \sin\phi_2=\frac{{\cal P}_2-1/2}{\sqrt{{\cal P}_3(1-{\cal P}_3)}}.
\end{equation}
In order to derive the general formula for the superposition of two pure nonorthogonal states of the qubit expressed in terms of spinors
\begin{equation}\label{S8}
|\chi_1\rangle=\left(\begin{array}{c}
a_1\\a_2\end{array}\right),\quad |\chi_2\rangle=\left(\begin{array}{c}
b_1\\b_2\end{array}\right)
\end{equation}
such that $|a_1|^2+|a_2|^2=|b_1|^2+|b_2|^2=1$ and defined as
\begin{equation}\label{S9}
|\chi\rangle=c_1|\chi_1\rangle+c_2|\chi_2\rangle
\end{equation}
we consider the matrix
\begin{equation}\label{S10}
|\chi\rangle\langle\chi|=\left(\begin{array}{c}
c_1a_1+c_2b_1\\
c_1a_2+c_2b_2\end{array}\right)\left(c_1^\ast a_1^\ast+c_2^\ast b_1^\ast,\,c_1^\ast a_2^\ast+c_2^\ast b_2^\ast\right).
\end{equation}
It has the form
\begin{equation}\label{S11}
|\chi\rangle\langle\chi|=\left(\begin{array}{cc}
|c_1a_1+c_2b_1|^2&(c_1a_1+c_2b_1)(c_1^\ast a_2^\ast+c_2^\ast b_2^\ast)\\
(c_1^\ast a_1^\ast+c_2^\ast b_1^\ast)(c_1a_2+c_2b_2)&|c_1a_2+c_2b_2|^2
\end{array}\right).
\end{equation}
The obtained matrix provides the density matrix of the superposition  state $\rho_\chi$ if one takes into account the normalization condition, i.e. $T=\langle\chi|\chi\rangle=\mbox{Tr}|\chi\rangle\langle\chi|$ providing relation $\mbox{Tr}\rho_\chi=1.$ Thus the density matrix is defined as $\rho_\chi=|\chi\rangle\langle\chi|/\langle\chi|\chi\rangle$
%
%
\begin{equation}\label{S12a}
\rho_{\chi}=\frac{1}{|c_1a_1+c_2b_1|^2+|c_1a_2+c_2b_2|^2}\left(\begin{array}{cc}|c_1a_1+c_2b_1|^2
&(c_1a_1+c_2b_1)(c_1^\ast a_2^\ast+c_2^\ast b_2^\ast)\\
(c_1^\ast a_1^\ast+c_2^\ast b_1^\ast)(c_1 a_2+c_2 b_2)
&|c_1a_2+c_2b_2|^2\end{array}\right).
\end{equation}
The superposition state with density matrix $\rho_\chi$ given by (\ref{S12a}) can be expressed in terms of three probabilities $P_1,\,P_2,\,P_3$ of the spin-$1/2$ projections $m=+1/2$ on the axes $x,\,y,\,z$, respectively, then
\begin{equation}\label{S13}
\rho_\chi=\left(\begin{array}{cc}P_3&P_1-1/2-i(P_2-1/2)\\
P_1-1/2+i(P_2-1/2)&1-P_3\end{array}\right).
\end{equation}
The density matrix (\ref{S13}) is associated with the normalized pure state $|\chi_0\rangle$ of the form
\begin{equation}\label{S14}
|\chi_0\rangle=\left(\begin{array}{c}
\sqrt{P_3}\\
\sqrt{1-P_3}e^{i\phi_s}
\end{array}\right),
\end{equation}
i.e. $\rho_\chi=|\chi_0\rangle\langle\chi_0|$. Here the phase $\phi_s$ is determined by the probabilities as
\begin{equation}\label{S15}
\cos\phi_s=\frac{P_1-1/2}{\sqrt{P_3(1-P_3)}},\quad \sin\phi_s=\frac{P_2-1/2}{\sqrt{P_3(1-P_3)}}.
\end{equation}
The relations (\ref{S15}) means that the mathematical formula where complex number $z=x+i y=|z|e^{i\phi(z)}$ has the probabilistic interpretation. Namely, for $|z|\leq1,\,|z|=\sqrt{p_3}$ the real and imaginary numbers of the phase factor $e^{i\phi(z)}=\cos(\phi(z))+i\sin(\phi(z))$ can be given in terms of the probabilities $0\leq p_1,p_2,p_3\leq1$ such that
\begin{eqnarray}
&&(p_1-1/2)^2+(p_2-1/2)^2=p_3(1-p_3),\quad ( p_2-1/2)^2+(p_3-1/2)^2=p_1(1-p_1),\nonumber\\
&&(p_3-1/2)^2+(p_1-1/2)^2=p_2(1-p_2). \label{R1}
\end{eqnarray}
The trigonometric functions can be always presented as functions of the probabilities, i.e.
\begin{equation}\label{R3}
\cos(\phi(z))=\frac{p_1-1/2}{\sqrt{p_3(1-p_3)}},\quad\sin(\phi(z))=\frac{p_2-1/2}{\sqrt{p_3(1-p_3)}}.
\end{equation}
Thus the set of complex numbers $z$ with property $|z|\leq1$ can be mapped onto the set of probability distributions describing the positions of three classical coins $(p_1,(1-p_1)),$ $(p_2,(1-p_2)),$ $(p_3,(1-p_3)),$ satisfying  (\ref{R1}). Formally the relation (\ref{R1}) corresponds to properties of the parameters describing the spinors which define the pure states of qubits. We will apply this interpretation to obtain the probability representation of matrix elements of arbitrary qudit density matrix.

Our aim is to express the probabilities $P_1,\,P_2,\,P_3$ as functions by probabilities $p_1,$ $p_2,$ $p_3,$ ${\cal P}_1,$ ${\cal P}_2,$ ${\cal P}_3$ and coefficients $c_1$ and $c_2$.
In order to have the expression of these probabilities $P_1,\,P_2,\,P_3$ as functions of arguments which are also the probabilities we introduce the following relations of the complex numbers $c_1,\,c_2$ with the formal probabilities $0\leq \Pi_1,\,\Pi_2,\,\Pi_3\leq1$ given by formulas for the projector $|\psi_c\rangle\langle\psi_c|$ where the spinor $|\psi_c\rangle$ reads
\begin{equation}\label{S16}
|\psi_c\rangle=\left(\begin{array}{c}
\sqrt{\Pi_3}\\
\sqrt{1-\Pi_3}e^{i\alpha}
\end{array}\right),
\end{equation}
and
\begin{equation}\label{S17}
\cos\alpha=\frac{\Pi_1-1/2}{\sqrt{\Pi_3(1-\Pi_3)}},\quad \sin\alpha=\frac{\Pi_2-1/2}{\sqrt{\Pi_3(1-\Pi_3)}}.
\end{equation}
We use the property that the phase of the complex coefficient $c_1$ can be chosen as zero as well as normalisation condition $|c_1|^2+|c_2|^2=1$ is compatible with the final expression (\ref{S12a}) of the density matrix of the superposition state. Comparing the density matrix (\ref{S12a}) with its expression in terms of probabilities $P_1,\,P_2,\,P_3$ we have the formulas
\begin{eqnarray}
&&P_3=\frac{|c_1a_1+c_2b_1|^2}{|c_1a_1+c_2b_1|^2+|c_1a_2+c_2b_2|^2},\nonumber\\
&&P_1-1/2-i(P_2-1/2)=\frac{(c_1a_1+c_2b_1)(c_1^\ast a_2^\ast+c_2^\ast b_2^\ast)}{|c_1a_1+c_2b_1|^2+|c_1a_2+c_2b_2|^2}.\label{S18}
\end{eqnarray}
Here we have to express the complex numbers $a_1,\,b_1,\,a_2,\,b_2,\,c_1,\,c_2$ in terms of probabilities, i.e., $a_1=\sqrt{p_3},\,b_1=\sqrt{{\cal P}_3},\,c_1=\sqrt{\Pi_3},\,a_2=\sqrt{1-p_3}e^{i\phi_1},\,b_2=\sqrt{1-{\cal P}_3}e^{i\phi_2},\,c_2=\sqrt{1-\Pi_3}e^{i\alpha}$, where phases $\phi_1,\,\phi_2,\,\alpha$ are given by Eqs.(\ref{S5}), (\ref{S7}), (\ref{S17}), respectively. The denominator in expression (\ref{S18}) reads
\begin{eqnarray}
&&{\cal T}=1+\frac{2}{\sqrt{p_3{\cal P}_3}}\left\{(\Pi_1-1/2)\left[(p_1-1/2)({\cal P}_1-1/2)+({\cal P}_2-1/2)(p_2-1/2)+p_3{\cal P}_3\right]+\right.\nonumber\\
&&\left.(\Pi_2-1/2)\left[(p_2-1/2)({\cal P}_1-1/2)-(p_1-1/2)({\cal P}_2-1/2)\right]\right\}.\label{S19}
\end{eqnarray}
The parameter ${\cal T}$ equals 1 if
\begin{equation}\label{eq.S19a}
\tan \alpha=\frac{1}{\sin (\phi_2-\phi_1)}\left\{ \frac {p_3{\cal P}_3}{\sqrt{p_3(1-p_3){\cal P}_3(1-{\cal P}_3)}}+\cos(\phi_1-\phi_2)\right\}.
\end{equation}
Then
\begin{equation}\label{S20}
P_3=\frac{1}{{\cal T}}\left\{\Pi_3p_3+(1-\Pi_3){\cal P}_3+2\sqrt{p_3{\cal P}_3}\left(\Pi_1-1/2\right)\right\}.
\end{equation}
\begin{eqnarray}
P_1-1/2=&&\frac{1}{{\cal T}}\left\{\Pi_3(p_1-1/2)+({\cal P}_1-1/2)(1-\Pi_3)+\right.\nonumber\\&&\left[(\Pi_1-1/2)(p_1-1/2)+(\Pi_2-1/2)(p_2-1/2)\right]\sqrt{\frac{{\cal P}_3}{p_3}}+\nonumber\\
&&\left.\left[(\Pi_1-1/2)({\cal P}_1-1/2)-(\Pi_2-1/2)({\cal P}_2-1/2)\right]\sqrt{\frac{p_3}{{\cal P}_3}}\right\}\label{S21}
\end{eqnarray}
\begin{eqnarray}
P_2-1/2=&&\frac{1}{{\cal T}}\left\{\left[(p_2-1/2)\Pi_3+({\cal P}_2-1/2)(1-\Pi_3)\right]+\right.\nonumber\\
&&\sqrt{\frac{{\cal P}_3}{p_3}}\left[(\Pi_1-1/2)(p_2-1/2)-(\Pi_2-1/2)(p_1-1/2)\right]+\nonumber\\
&&\left.\sqrt{\frac{p_3}{{\cal P}_3}}\left[(\Pi_2-1/2)({\cal P}_1-1/2)+(\Pi_1-1/2)({\cal P}_2-1/2)\right]
\right\}.\label{S22}
\end{eqnarray}
As we have shown in explicit form the probabilities determining the superposition states in the both cases where $\langle\psi_1|\psi_2\rangle=0$ and $\langle\psi_1|\psi_2\rangle\neq0$ are expressed in terms of the probabilities determining the states $|\psi_1\rangle$ and $|\psi_2\rangle.$

\section{Conclusion}
To resume we formulate the main result of our study. We have shown that the state of qubit can be given as the set of three probability distributions $(p_1,1-p_1)$, $(p_2,1-p_2)$, $(p_3,1-p_3)$. The superposition principle of the pure qubit states is presented in the form of the addition rule of these probability distributions. The relative phase of the superposition responsible for interference phenomenon is also expressed in terms of the probability distributions. The addition rule for the probabilities can be illustrated as the rule of combination of two triadas of Malevich's squares. Our main results are formulas (\ref{S19}), (\ref{S20})-(\ref{S22}). The extension of the obtained nonlinear addition rule for the probabilities determining the superpositions of arbitrary qudit states will be given in future publications. One can illustrate the obtained map of the quantum states to the classical coins probabilities by using the following statement of A. Einstein "God does not play dice" and replacing it by statement "God plays coins".


\begin{thebibliography}{99}
\bibitem{Dirac}P.~Dirac, {\it The Principles of Quantum Mechanics}, Oxford University Press (1930).
\bibitem{Sch26} E.~ Schr\"odinger, {\sl Naturwissenchaften}, {\bf Bd.~14}, s.~664 (1926).
\bibitem{Landau27} L.~D.~Landau, {\sl Z. Physik}, {\bf 45}, p.~430 (1927)
\bibitem{vonNeumann27} J.~von~Neumann, {\it Mathematische Grundlagen der Quantenmechanik}, Springer, Berlin (1932).
\bibitem{Sudarshan1}V. I. Man'ko, G. Marmo, E. C. G. Sudarshan, F. Zaccaria, {\sl J. Phys.A: Math. and Gen.}, {\bf 35}, 7137 (2002).
\bibitem{Sudarshan2}V. I. Man'ko, G. Marmo, E. C. G. Sudarshan, F. Zaccaria, {\sl Phys. Lett. A},{\bf 327}, 353 (2004)
\bibitem{TombesiPLA}S.~Mancini, V.~I.~Man'ko, and P.~Tombesi, {\sl Phys.~Lett.~A}, {\bf 213}, 1 (1996).
\bibitem{DodPLA}V.~V.~Dodonov and V.~I.~Man'ko, {\sl Phys.~Lett.~A}, {\bf 239}, 335 (1997).
\bibitem{OlgaJETP}V.~ I.~Man'ko and O.~V.~Man'ko, {\sl J. Exp. Theor. Phys.}, {\bf 85}, 430 (1997).
\bibitem{Bregence}O.~V.~Man'ko, in: B.~Gruber and M.~Ramek~(Eds.), {\it Proceedings of International Conference ``Symmetries in Science X'' (Bregenz, Austria, 1997)}, Plenum Press, New York (1998), p.~207.
\bibitem{Weigert1}S.~Weigert, {\sl Phys.~Rev.~Lett.}, {\bf 84}, 802 (2000).
\bibitem{Weigert2} J.-P.~Amiet and S.~Weigert, {\sl J.~Opt.~B: Quantum Semiclass. Opt.}, {\bf 1}, L5 (1999).
\bibitem{Paini} G.~M.~D'Ariano, L.~Maccone, and M.~Paini, {\sl J.~Opt.~B: Quantum Semiclass. Opt.}, {\bf  5}, 77 (2003).
\bibitem{IbortPhysScr150}A.~Ibort, V.~I.~Man'ko, G.~Marmo, A.~Simoni, and F.~Ventriglia, {\sl Phys. Scr.}, {\bf 79},
    065013 (2009).
\bibitem{Marmo}Vladimir~I.~Man'ko, Giuseppe~Marmo, Franco~Ventriglia, and Patrizia Vitale, {\sl J. Phys. A: Math. Gen.}, {\bf 50}, 335302 (2017); arXiv:1612.07986 (2016).
\bibitem{Chernega1}V.~N.~Chernega, O.~V.~Man'ko, and V.~I.~Man'ko, {\sl J. Russ. Laser Res.}, {\bf 38},
141 (2017).
\bibitem{Chernega2}V.~N.~Chernega, O.~V.~Man'ko, and V.~I.~Man'ko, {\sl J. Russ. Laser Res.}, {\bf 38},
324 (2017).
\bibitem{Chernega3}V.~N.~Chernega, O.~V.~Man'ko, and V.~I.~Man'ko, {\sl J. Russ. Laser Res.}, {\bf 38},
416 (2017).

\end{thebibliography}
\end{document}